\documentclass{aa}  
\usepackage{graphicx}
\usepackage{txfonts}
\def\kms{kms$^{\rm -1}$}
\def\vmic{$v_{\rm mic}$}

\def\wzup{\omega_{\rm z}^{\rm up}}
\def\wzdown{\omega_{\rm z}^{\rm down}}
\def\ww{\overline{\omega^2}}

\def\www{\overline{\omega^3}}

\def\logtauross{{\rm log(}\tau_{\rm ross}{\rm )}}
\begin{document}
   \title{ Modelling convection in A star atmospheres }

   \subtitle{ Bisectors and lineshapes of HD\,108642 }

   \author{ Ch.\,St\"utz \inst{1} }

   \offprints{Ch.\,St\"utz}

   \institute{Institute for Astronomy (IfA), University of Vienna,
              T\"urkenschanzstrasse 17, A-1180 Vienna\\
             }

   \date{Received ; accepted }

  \abstract
   { We present a code, VeDyn, for modelling envelopes and atmospheres of A to F stars
     focusing on accurate treatment of convective processes.
   }
   { VeDyn implements the highly sophisticated convection model of 
     Canuto and Dubovikov (1998) but is fast and handy enough to be 
     used in practical astrophysical applications.
   }
   { We developed the HME envelope solver for this convection model furtheron
     to consistently model the envelope together with the stellar atmosphere.
     The synthesis code SynthV was extended to account for the resulting 
     velocity structure. Finally, we tested our approach on atomic line bisectors.
   }
   { It is shown that our synthetic line bisectors of HD\,108642 bend towards the blue 
     and are of a magnitude comparable to the observed ones.
   }
   { Even though this approach of modelling convection requires the solution of a coupled
     system of nonlinear differential equations, it is fast enough to be applicable to 
     many of the investigation techniques relying on model atmospheres.
   }

   \keywords{ 
              Convection -- Turbulence -- Stars: atmospheres -- Line: profiles -- Stars individual: HD108642
            }

   \maketitle


\section{Introduction}
A physical mechanism most challenging to understand in F to early A stars is convection.
It is one of the least understood phenomena taking place in these stars.
Recently, several successful attempts to describe and to model convection in 
the envelope and the photosphere in more detail have been conducted. 
To be mentioned in this regard are for instance,
numerical simulations by
Nordlund \& Stein (\cite{ns2000}), Stein \& Nordlund (\cite{sn2003}),
Freytag \& Steffen (\cite{freytag04}) and Kochukhov et al. (\cite{kochuk07}),
but also tests and applications of nonlocal convection models presented in
Kupka (\cite{k1999}, HME) , Kupka \& Muthsam (\cite{km2007}),
and Kupka \& Montgomery (\cite{km2002}).

However, in practical applications like stellar spectrum analysis or
stellar evolution modelling the mixing length theory (Biermann \cite{mlt1948})
is still most commonly used to model convection.
Kupka (\cite{k1996}) and Heiter et al. (\cite{nemo2002}) implemented a 
local full spectrum turbulence model of Canuto and  Mazzitelli (\cite{cm1991},\cite{cm1992})
and Canuto, Goldman and Mazzitelli (\cite{cgm1996}) in the 
widely used model atmosphere code 
{\sc Atlas9} (Kurucz \cite{atlas9}).
The complete Reynolds stress model (RSM) of Canuto and Dubovikov (\cite{cd1998}, CD98) 
has been applied to envelopes of A--stars and also white dwarfs a few years 
ago (Kupka \& Montgomery \cite{km2002}, Montgomery \& Kupka \cite{mk2004}).
Their calculations are in qualitative agreement with 
2D numerical simulations of Freytag (\cite{freytag95}) and the resulting 
vertical velocites are consistent with observerved values for micro-- and 
macroturbulence.

Extending this work, we investigated the possibility to apply the CD98 convection
model also to stellar atmospheres. The various analysis methods of stellar spectra
connected to or relying on stellar atmosphere modelling mostly use the micro- and
macroturbulence parameters to account for those contributions to the lines shapes 
believed to originate from turbulent velocities. These however, permit only to 
model distortions symmetric with respect to the line center and are not inherent
to the underlying stellar model.

In this publication we show our extension of the HME envelope solver to model
stellar envelopes together with the stellar atmosphere and a method to account for 
the derived velocity fields in spectrum synthesis. Following this approach, the 
structure and magnitude of the up- and downflows is resulting from the model
calculation and is then put into the derivation of spectral line profiles.
Hence, the need for the parametrization of a micro- and marcroturbulence 
is expected to be considerably reduced.

We also present an application of our models to the star HD\,108642 
which has already been investigated by Landstreet (\cite{land1998}) and others concerning
microturbulence, abundances and also bisectors.


\section{The models}

The convection model which we used is an extension of the CD98 approach, as described in
Kupka and Montgomery (\cite{km2002}). The equations for the turbulent quantities
are solved on a weighted mass grid. The weighting allows a non--equidistant spacing of
the grid such that steep temperature gradients are resolved. All other quantities are
derived on the same weighted mass grid in the envelope and on a log($\tau_{\rm ross}$) 
grid in the atmosphere.
The transition region is determined automatically according to $\rho$, $P$ and $T$ at each 
iteration step. This way we could consistently model the stellar atmosphere together 
with the envelope with feasible computational efforts. 
For the models presented in this paper we assumed spherical symmetry.
We aim to use at least two sources for the opacities which are {\sc{OPAL}} 
opacities (\cite{opal1996}) in the envelope, and opacitiy distribution functions in 
the atmosphere. At present, however, we restrict ourselves to the {\sc OPAL}  routines, 
since they are faster and basic code development is still ongoing.  
Depending on the maximum size of time steps and the necessary relaxation time one model
requires 3 to 30 hours to finish on a 2.3\,GHz single CPU machine. 

\section{Synthesis and bisectors}

To derive synthetic spectra from our models and velocity profiles, we
adopted the SynthV code of Vadim Tsymbal.
Following the CD98 equations (38d) and (50d) to (50f)
\begin{eqnarray}
  S_w     & = & \frac{\www}{(\ww)^{3/2}} \\
  \sigma  & = & 0.5 \, [ 1 - S_w\,(4+S_w^2)^{-1/2} ] \\
  \wzup   & = & - (\frac{1-\sigma}{\sigma})^{1/2} \, \ww^{1/2} \\
  \wzdown & = &   (\frac{\sigma}{1-\sigma})^{1/2} \, \ww^{1/2}
\end{eqnarray}
we can derive typical velocities $\wzup$, $\wzdown$ and the filling factor $\sigma$
(fractional area occupied by the updrafts)
from the mean vertical turbulent velocity $\ww$ and the skewness $S_w$.
\begin{figure}[h]
\begin{center}
\includegraphics[height=88mm, angle=270]{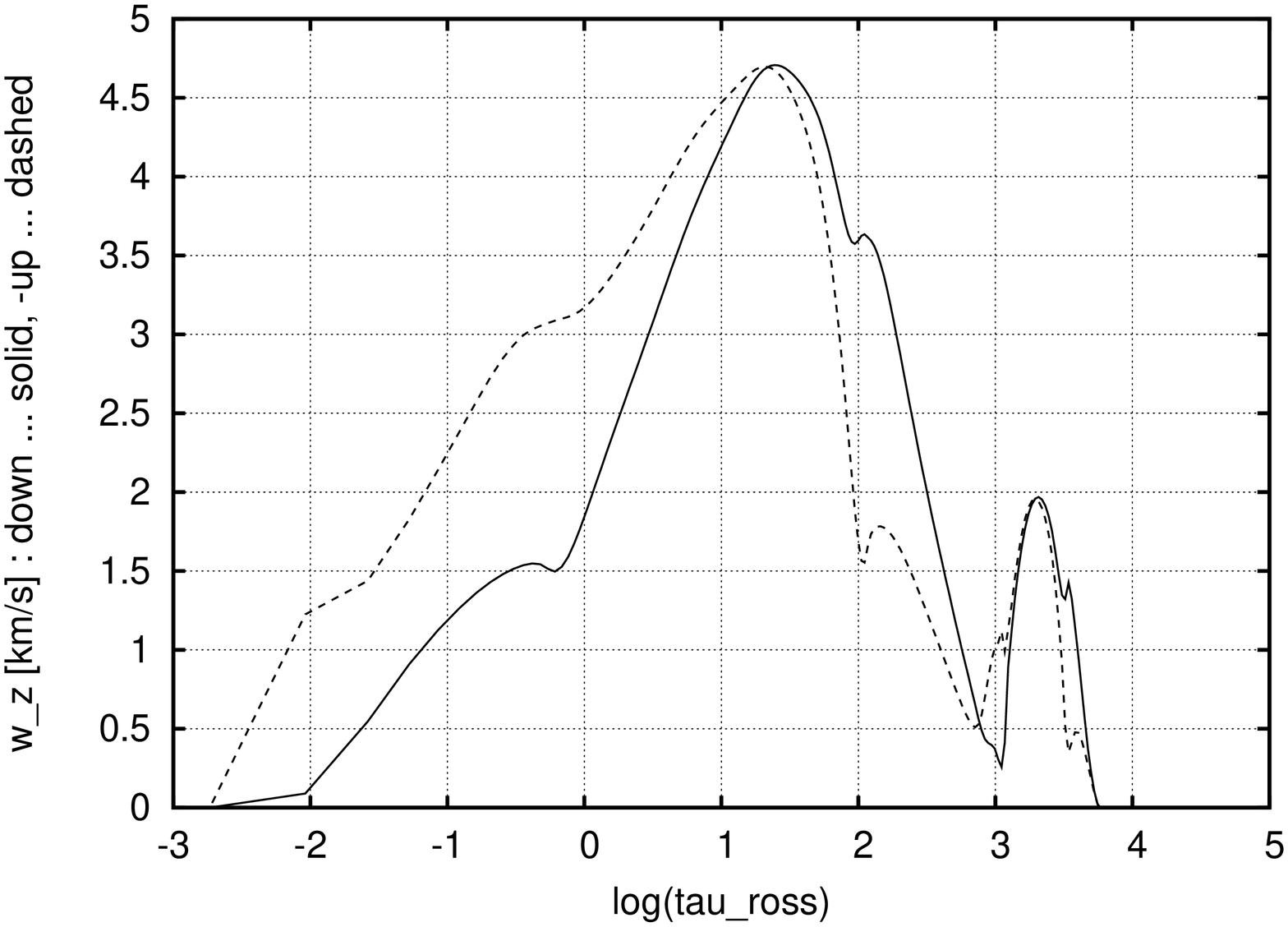}
\caption{Velocities ($-\wzup$ - dashed, $\wzdown$ - solid) for HD\,108642
         derived from our model.} 
\label{velocity}
\end{center}
\end{figure}

Figure \ref{velocity} compares the absolut values of the up- and downdraft velocities we
calculated. Note that at tau = 2/3 ($\logtauross$\,=\,-0.176) these are in the
range of typical values for the microturbulent velocity one would expect for an
A type star. Contrary to CD98 our upstream velocities are negative, since in our
frame of reference up means moving towards the observer.

\begin{figure}[h]
\begin{center}
\includegraphics[height=88mm, angle=270]{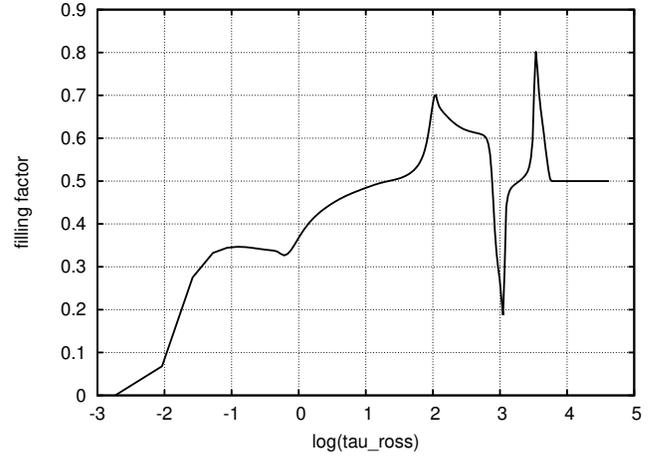}
\caption{Filling factor in our model of HD\,108642.}
\label{filling}
\end{center}
\end{figure}

The synthesis code SynthV of V.\,Tsymbal has been developed further on to
allow to account for the upward and downward flow by means of separate radial velocity 
profiles. Their contributions to the total line profile are then summed up 
weighted according to the filling factor (Figure \ref{filling}). \\

\section{HD\,108642}

The Am star HD\,10842 has been observed several times over the past years
concerning its peculiar nature, abundance pattern and also (Landstreet \cite{land1998})
atmospheric velocity fields. 
The observations we used for this investigation were done by Landstreet and Kupka in 
April 2001 at the CFHT. They observed the star at a very high resolution of R = 120000
in several chunks of approximately 100\,{\AA} in range. We found the ranges  5491 - 5572\,{\AA}
and 6102 - 6191\,{\AA} most suitable. These contain enough unblended lines at a reasonable 
noise level S/N $>$ 140 and 150.

As starting point for our model for the envelope and atmosphere of HD\,108642 we took 
evolutionary envelopes as well as polytropes. We found our code to be insensitive to
the differences of these types of envelope models.
Extension into the atmosphere was performed with the gray approximation. 
For radiation transport we used the OPAL opacities and equation of state 
Rogers et al. \cite{opal1996}) as input data.
The only tuneable parameters we had in our model of HD\,108642 were its effective
temperature of 8100\,K, the stars surface gravity of 4.1 and we accounted 
for its higher metallicity by using opal tables for Z = 0.06.
All other parameters and numerical constants are as described in 
Kupka and Montgomery (\cite{km2002}).
To do the line synthesis we adopted the SynthV code of Tsymbal. Its approach of solving
the equation of radiative transfer for each atmospheric layer separately naturally 
allows to account for depth dependent velocity fields. We performed the radiative transfer 
twice for each layer, for the upwards and for the downwards moving matter. These two 
contributions were then summed up weighted according to the filling factor and the absorption
of the individual layer added to the total line absorbion. 

\begin{figure}[h]
\begin{center}
\includegraphics[height=88mm, angle=270]{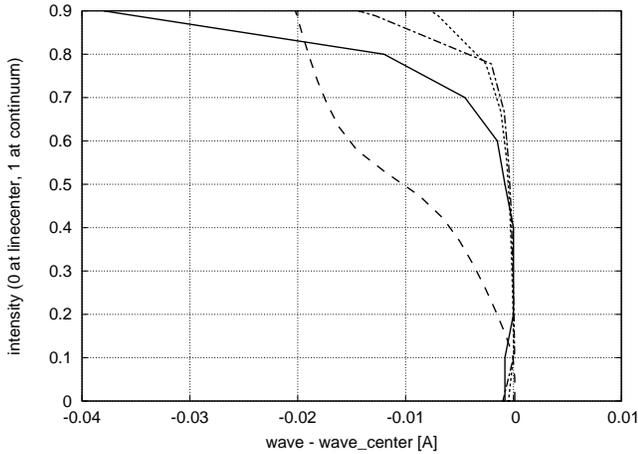}
\caption{The solid line shows the observed bisector of Cr{\sc II} 
         as measured by Landstreet 1998, the dashed line denotes the 
         synthesized bisector. 
         The dotdashed and dotted line represent the bisectors of Co{\sc{I}}\,5342.7
         and Fe{\sc{I}}\,5241.9 respectively; both weak lines.
        } 
\label{bisector}
\end{center}
\end{figure}

Figure\,\ref{velocity} shows the mean velocities of upwards and downwards moving 
material resulting from our calculations. The values of $\wzup$ and $\wzdown$ taken
at $\tau_{ross}$ = 2/3 compare reasonably to the measurement of 
\vmic = 4.0\,\kms by Landstreet (\cite{land1998}).
A comparison of the observed bisector of the Cr{\sc II} line at 4616\,{\AA} 
(Landstreet \cite{land1998}) and the synthetic bisector derived from the VeDyn model 
can be seen in Figure\,\ref{bisector}. Clearly the magnitude of the calculated bisector is 
of the same order as the observed one. Also the tilt to the blue is reproduced.
However, its shape is quite deviating. This, we believe, is at least in part due to 
the fact that we could not include the upper part of the atmosphere 
(log($\tau_{ross}$) $<$ -4.0), because of numerical problems. 
Furthermore figure \ref{bisector} shows the bisectors of two weak lines,
Co{\sc{I}}\,5342.7 and Fe{\sc{I}}\,5241.9; an encouraging result of our modelling procedure.
However, we currently do not have any observations to compare those synthetic bisectors to.

\section{Conclusions}

We consistently modeled turbulent convection in the lower atmosphere and envelope of an Am star.
Following the model of Canuto and Dubovikov (\cite{cd1998}) and the approach of
Kupka \& Montgomery (\cite{km2002}) we solved the equations for the turbulent quantities 
on a weighted mass grid. Other quantities are computed on the same weighted mass grid
in the envelope and on a $\logtauross$ grid in the stellar atmosphere.
Line synthesis has been performed with a modified version of the SynthV code of Vadim Tsymbal,
where we account for the depth dependent mean up- and downstream velocities and the
filling factor.

The resulting line shapes of HD\,108642 have been compared to the observed ones. 
The same has been done
for the typical turbulent velocities we derive at optical depths in the vicinity of 
$\tau_{ross}$ = 2/3. They are in the range of the microturbulent velocity measured by
Landstreet (\cite{land1998}).

We also calculated bisectors from our synthetic lineprofiles. They are tilted to the blue 
by a magnitude comparable to the observed ones. However, the shape of the synthetic
bisector differs from the observation. We think this results from the upper part of 
the atmosphere missing in our model. 
Currently more objects are investigated to cover the region of A to F type main sequence stars.

Our modelling and synthesis approach is applicable to investigation methods
relying on model atmospheres. Still, the frequency dependent radiation transport has to be 
implemented in our models.

\begin{acknowledgements}

The author likes to thank Vadim Tsymbal for letting us alter his synthesis code SynthV, 
F.\,Kupka who developed the HME solver for his support and J.\,Landstreet for the 
observational data on HD\,108642. 
This research was funded by the FWF project P-18224-N13.

\end{acknowledgements}


\end{document}